\documentclass[conference]{IEEEtran}
\IEEEoverridecommandlockouts
\usepackage{cite}
\usepackage{amsmath,amssymb,amsfonts}
\usepackage{algorithmic}
\usepackage{graphicx}
\usepackage{pdfpages}
\usepackage{textcomp}
\usepackage{xcolor}
\usepackage{diagbox}
\usepackage{url}
\usepackage{listings}
\usepackage{multicol}

\def\BibTeX{{\rm B\kern-.05em{\sc i\kern-.025em b}\kern-.08em
    T\kern-.1667em\lower.7ex\hbox{E}\kern-.125emX}}
\begin{document}

\title{Understand Code Style: Efficient CNN-based Compiler Optimization Recognition System
}

\author{\IEEEauthorblockN{ Shouguo Yang\IEEEauthorrefmark{1}\IEEEauthorrefmark{2}, Zhiqiang Shi\IEEEauthorrefmark{1}\IEEEauthorrefmark{2}, Guodong Zhang\IEEEauthorrefmark{1}\IEEEauthorrefmark{2}  
\\Mingxuan Li\IEEEauthorrefmark{1}\IEEEauthorrefmark{2},
  Yuan Ma\IEEEauthorrefmark{1}\IEEEauthorrefmark{2}, Limin Sun\IEEEauthorrefmark{1}\IEEEauthorrefmark{2}}
\IEEEauthorblockA{\IEEEauthorrefmark{1}\textit{Institute of Information Engineering, Chinese Academy of Sciences, Beijing, China} \\
\IEEEauthorrefmark{2}\textit{School of Cyber Security, University of Chinese Academy of Sciences, Beijing China}\\
\{yangshouguo, shizhiqiang, zhangguodong, limingxuan, mayuan2, sunlimin\}@iie.ac.cn}
}

\maketitle

\begin{abstract}
 Compiler optimization level recognition can be applied to vulnerability discovery and binary analysis. Due to the exists of many different compilation optimization options, the difference in the contents of the binary file is very complicated. There are thousands of compiler optimization algorithms and multiple different processor architectures, so it is very difficult to manually analyze binary files and recognize its compiler optimization level with rules. This paper first proposes a CNN-based compiler optimization level recognition model: BinEye. The system extracts semantic and structural differences and automatically recognize the compiler optimization levels. The model is designed to be very suitable for binary file processing and is easy to understand. We built a dataset containing 80028 binary files for the model training and testing. Our proposed model achieves an accuracy of over 97\%. At the same time, BinEye is a fully CNN-based system and it has a faster forward calculation speed, at least 8 times faster than the normal RNN-based model. Through our analysis of the model output, we successfully found the difference in assembly codes caused by the different compiler optimization level. This means that the model we proposed is interpretable. Based on our model, we propose a method to analyze the code differences caused by different compiler optimization levels, which has great guiding significance for analyzing closed source compilers and binary security analysis.

\end{abstract}


\section{Introduction}
With the scale of IoT devices getting larger and larger, the security of software systems running on IoT devices is becoming more and more important.
IoT software security is increasingly affected by different compiler optimization levels.
With using different compilation optimization levels, the same source code can generate different binary code. The binary code can be different in terms of function control flow, data flow, function inlining style, and so on.
These differences can easily lead to security risks.
According to the previous work, compiler optimization levels recognition can be used in areas such as vulnerability discovery\cite{b2} and binary similarity detection\cite{b3}.
Different compiler optimization levels have a significant impact on software security\cite{b1}. Compiler optimization levels are an important reference for software security analysis.
For example, the compiler optimization algorithm "dead store elimination" causes a very famous security problem CWE-14\cite{b4}.
The "undefined behavior" which means compilers are free to decide how to handle code optimization in C/C++ can also create security vulnerabilities in some cases such as CVE-2016-9843 and CVE-2016-9840.

Today's common compilers, such as GCC, have thousands of compiler optimization algorithms built in, each of which corresponds to an option at compile time. 
For ease of use, compilers such as GCC integrate numerous compiler optimization levels into five broad levels -O0, -O1, -O2, -O3 and -Os.
The smallest optimization-associated unit generated by compilers is the object file. The object files compiled with different compilation optimization levels may be linked to the same executable binary file, which we can not identify the compiler optimization of the executable binary file directly, so the system target file we aim for is the compiler-generated object file. But a variety of compiler optimizations options causing a tiny difference in the binary code of object file, which is very time-consuming with the human to observe, and need to be very rich experience in assembly language. So we propose an end-to-end deep learning model to recognize different compiler optimization levels of an object file. In order to be more easily understood, we successfully build the model without any compiler knowledge and analysis for the binary file, which means it is easy to handle security problems causing by compiler optimization.

   With reference to Facebook's great success in using CNN in machine translation applications in the field of language processing\cite{b8} and the application of CNN in malware detection\cite{b7}, the performance of CNN in language processing and binary security analysis is not inferior\cite{b8}. In addition, CNN can make full use of GPU for parallel computing and more efficient parameter gradient calculation and training compared with RNN.  So, our model is fully based on CNN technology, which can be used to identify compiler optimization levels directly from the raw data in the executable binary. 
According to previous work\cite{b6}, the binaries compiled from compilation optimization level -O2 or -O3  are not very different.
Since the compiler optimization levels -O2 and -O3  are almost identical for the compilation of most source code, we define a four-category task: identify -O0, -O1, -O2/-O3, -Os four different compiler optimization levels from the binary file. We perform a lot of experiments with our model for the four-category task.
 The experimental results show that the accuracy of our model for the four-category task can reach 97\%. Compared with the model which is aimed at the binary files on the X64 platform proposed by Chen\cite{b9}, Our model is aimed at the binary files under the ARM platform. Since our model directly takes the raw binary bytes as input, the model does not require any prior knowledge and complex feature engineering. It can be trained and converge more quickly. The model is smaller and the forward calculation speed is also faster. 
 We successfully found several assembly code differences caused by different levels of compiler optimization through the output of the model, and summarized a method for code analysis with the model.
Due to the characteristics of the RISC, the model can be easily extended to other compilers and instruction sets which belongs to other RISC platform.
 
   The contributions of this paper are summarized as follows:
\begin{itemize}
        \item We propose a model: \textbf{BinEye} to recognize compiler optimization levels, which does not require any prior knowledge and feature engineering. The object file is directly fed to our model as input. The model based on CNN can more fastly perform parallel training and testing, perform very fast forward calculations. 
    
    \item We built a dataset consisting of \textbf{80028} object files totaling 691MB. We perform experiments with our model and the model achieved a high recognition accuracy.
    
    \item We propose a method that can quickly and clearly analyze the differences caused by different compilation optimization levels with the help of the output of our model.

\end{itemize}

\section{Related Work}

Prior to this, there were a lot of researches on software security in terms of compiler optimization levels. D’Silva et al.\cite{b1} proposed a broad research programme whose goal is to identify, understand, and mitigate the impact of security errors introduced by different compiler optimizations. Wang et al.\cite{b2} proposed a novel model, which views unstable code in terms of various compiler optimizations that leverage undefined behavior. Different compiler optimizations which introduce unstable code may cause a security problem. Using their model, they had uncovered 160 new bugs that have been confirmed and fixed by developers. It can be learned that different compiler optimization levels could cause security vulnerabilities.

Due to the existence of many kinds of programming languages and CPU architectures in the real world, it makes the methods using rules to handle security problem more complicated, which also needs stronger professional expertise and capability, takes a long time to analyze and formulate rules. And according to the study of Chen el at.\cite{iot}, the process of static binary executable analysis involves lots of manual, repetitive efforts, like adjusting load offsets and handling disassembling errors. It is limited to get the semantic information due to difficulties in disassembly. So we try to capture semantic and structural patterns directly from the binary code. With the successful application of deep learning in the fields of image processing, language processing and so on, deep learning is also applied to the security field. The neural network can automatically learn and memorize complex patterns in programming languages through deep learning techniques. 
For example, the CNN-based malware recognition model proposed by Edward et al.\cite{b7} can better identify the maliciousness of the software. The model can automatically learn the characteristics of the binary code in the malware to distinguish malware and non-malware.  
Shihab\cite{Shihab} proposed an efficient and scalable technique for computer network security which using a multi-layer neural network for decryption scheme and public key creation.
Xu et al.'s\cite{b11} use of graph-encoded neural networks for vulnerability function match tasks has also achieved good results. Shin\cite{b12} propose to apply artificial neural network to solve important yet difficult problems in binary analysis, that is function recognition, which is a crucial first step in binary analysis techniques. The result shows that deep learning technique can identify functions in binaries better than other machine-learning-based methods. 
The deep learning model based on RNN proposed by Chen et al.\cite{b9} is also applied to compiler optimization levels recognition but requires strong expertise for feature engineering and complex data preprocessing, and his work only focus on X86 architecture. The feature extraction in their work from the assembly code need the disassembly process, as mentioned before, there are many disassembling errors in the disassembly process. So, the method proposed by Chen is not accurate enough. 

Inspired by the above work, we applied a new CNN-based neural network to compiler optimization levels recognition tasks and achieved good results. Compared with the previous work of Chen\cite{b9}, we improved the accuracy of the 4 classification tasks and simplified the compiler optimization levels recognition task: directly identify the compiler optimization level of object files without extracting features and complex preprocessing. The data fed into our model does not need the disassembling process which may introduce some flaws. The performance of our model is much higher than the model using RNN in terms of model size, forward calculation speed, etc.

\section{Detailed Design}

\begin{figure}[htbp]
    \centerline{\includegraphics[width=3.5in]{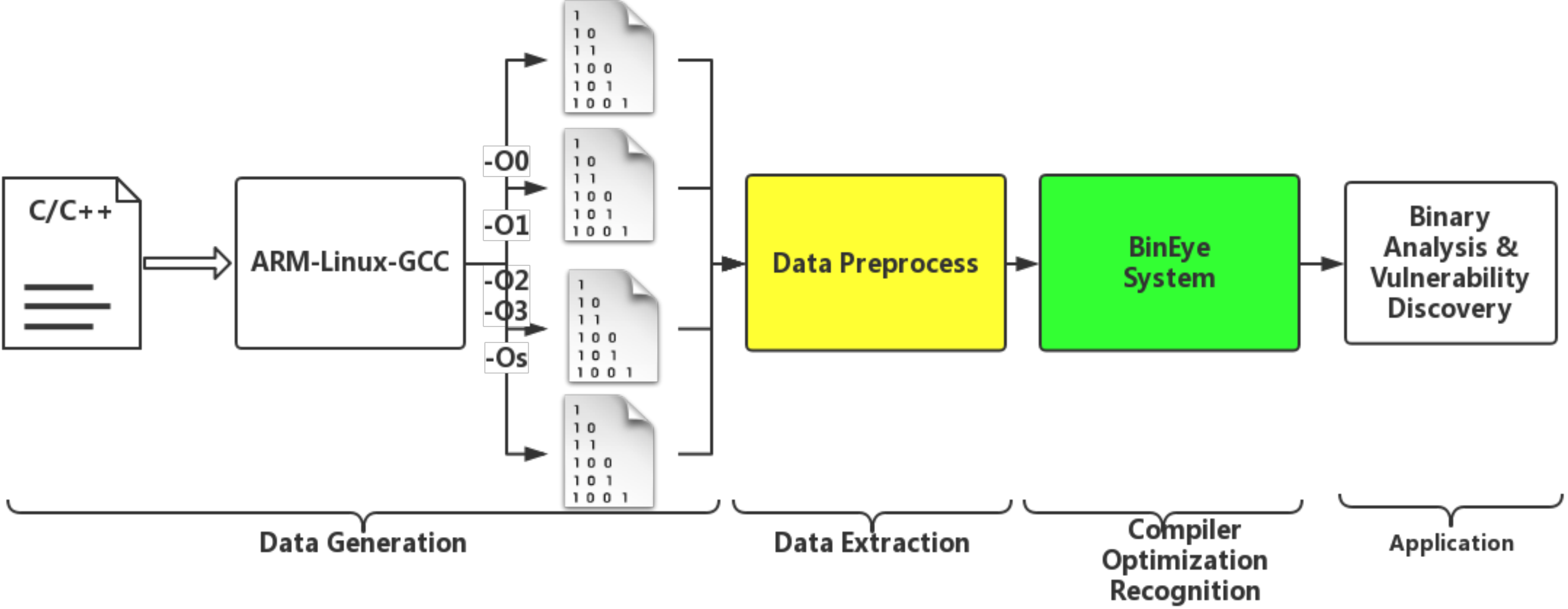}}
    \caption{System Architecture.}
    \label{fig1}
\end{figure}

The overall system architecture is shown in Fig. \ref{fig1}. 
First part is data generation part. In this part, we  compiled sources codes for five different compiler optimization levels with cross-compilation toolchain arm-linux-gnueabi-gcc. After compilation, we get four kinds of different object files -O0, -O1, -O2/-O3, -Os. Then we extract code segment data from object files as inputs of \textbf{BinEye}. Code segment binary data is fed into model and it performs the compiler optimization levels recognition task. With the intermediate output of the model, we can analyze the assembly code
,which we discussed in section \ref{analysis}.After we get the information about compiler optimization levels, we can do other mission such as binary analysis and vulnerability discovery, which we do not discuss in this paper.

\begin{table}[htbp]
\caption{Dataset Detail}
\begin{center}
\begin{tabular}{|c|c|c|}
\hline
\textbf{Compiler Optimization Level}&\textbf{Object files} & \textbf{total size(MB)}\\
\hline
\textbf{-O0} & 19076& 209 \\
\hline
\textbf{-O1} & 17136& 131 \\
\hline
\textbf{-O2} & 13768& 112 \\
\hline
\textbf{-O3} & 13992& 128 \\
\hline
\textbf{-Os} & 16056& 112 \\
\hline
\textbf{total} & \textbf{80028}& \textbf{691} \\
\hline
\end{tabular}
\label{tab1}
\end{center}
\end{table}

\begin{figure*}[htbp]
    \centerline{\includegraphics[width=11cm, height=14cm]{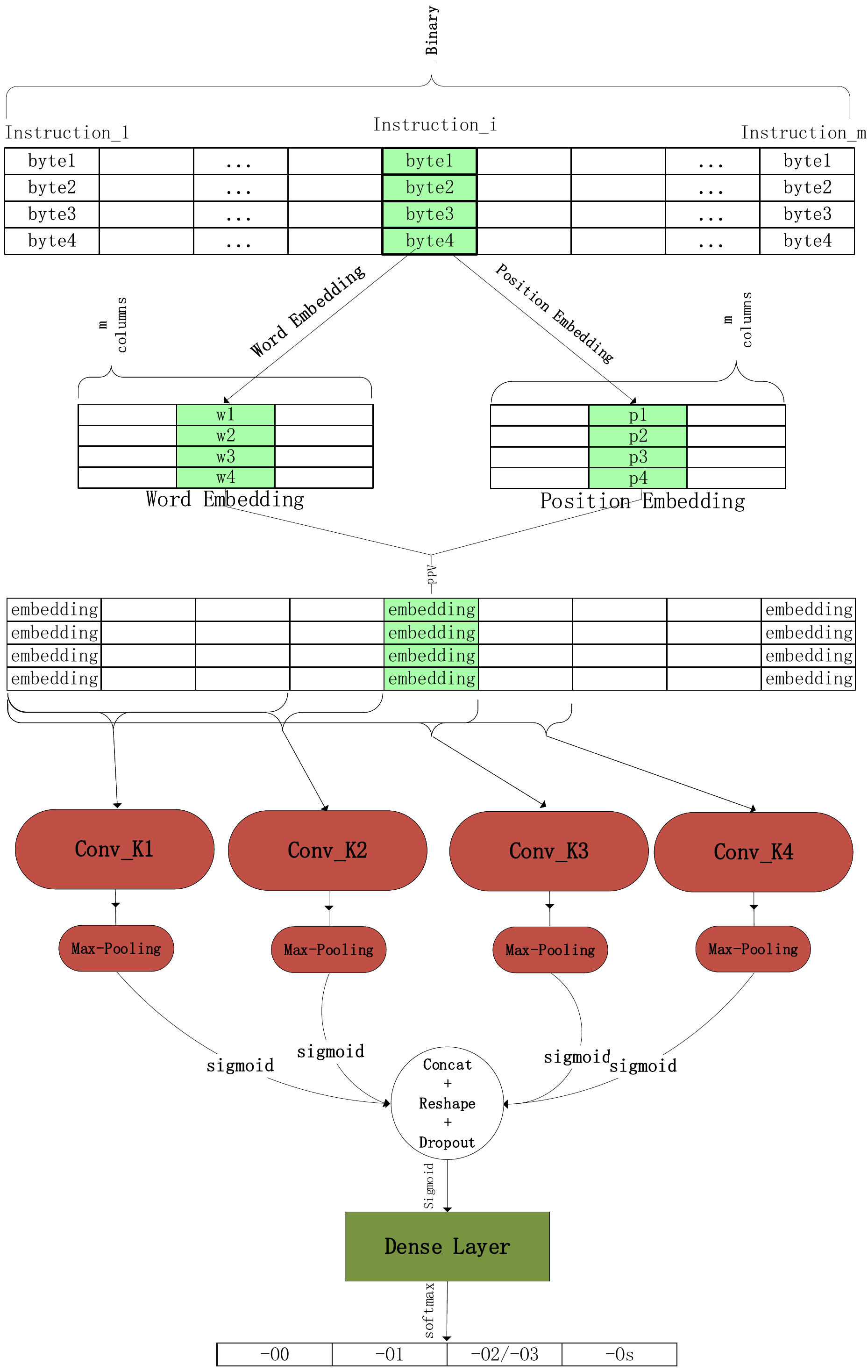}}
    \caption{BinEye Model Structure.}
    \label{fig2}
\end{figure*}

\subsection{Data Generation and Extraction}

We use buildroot\cite{buildroot} cross-compilation toolchain to compile 463 open source components with five different compiler optimization levels -O0,  -O1,  -O2,  -O3,  -Os,  and generated a total of 80028 object files (See Table \ref{tab1} for details).  The command to configure with buildroot is \textbf{"make menuconfig"},  and the command to download the component source code is "make source". To generate object files with different compiler optimization levels, we replace the string associated with the compilation optimization level configuration ("-O0,  -O1,  -O2,  -O3,  -Os") in the component source package by the shell command and replace them with the target optimization level. Then go back to the root directory of the buildroot and execute the "make" command to compile the source code according to the specified optimization level. Since the binary file is linked by the object file,  and the object file is the smallest compilation unit,  this work only considers the object file when constructing the dataset.  It should be noted that even if the compilation optimization level is specified in the previous step,  there is no guarantee that all source code will be successfully configured (Because compilation optimization level of some source code is not specified in the makefile). Therefore, we remove all MD5 signature-consistent files among compiled object file dataset by the principle of insufficiency,  so as to ensure that the selected binary files have the correct compilation optimization level.

According to the structure of the ELF file, many sections of the content are not affected by the compiler optimization levels. So, we only focus on the content of the code segment. It is easy to get the content of the code segment with the "readelf" tool. Since ARM architecture belongs to RISC, after we get 4-byte aligned binary data with the "readelf" tool, the binary data was directly fed into \textbf{BinEye} Model without any processing.

\subsection{Model Structure}

The model mainly consists of three parts: data re-presentation part, convolution and pooling part, and output part.
 The detailed design is shown in Fig. \ref{fig2}, Since the convolution and pooling parts are the more commonly used modules in CNN, we only show the data embedding representation part  \ref{we} and \ref{pe}, and our special structure in CNN \ref{cnn}. The various parts are introduced as followed sections.

\subsubsection{Word Embedding}

\label{we}
We extract $1024 \times 4$ bytes from the code segment of the binary object file as the original input to the neural network. Because the ARM architecture is a kind of RISC, each instruction occupies a fixed 4 bytes,  extracting $1024 \times 4$ bytes is equivalent to extracting 1024 instructions as neural network input. This method of data extraction can also be applied to the architecture of all RISC, such as MIPS, PowerPC, and so on. The first layer of the neural network is the word embedded layer, which is used to encode the input binary instructions bytes. We use the embedded layer to re-present each byte with a 4-dimensional vector in order to find and represent the semantic similarity of certain instructions.

In order to better capture the characteristics of RISC,  we embedded input $x = (x_1,...,x_i,...,x_m)$ in a distributed space as $w=(w_1,...,w_j...,w_m)$, where $w_j \in R^f $  is a column in an embedding matrix $D \in R^{f \times V}$ , Where V is the number of possible values of the input data. Since each byte has a maximum of $2^8$ values, so $V=2^8$, $f $ represents one byte from input represented by an f-dimensional vector.

In previous work \cite{Bengio}, we can know that word embedding has the advantage of data dimension reduction compared to one-hot data representation. On the other hand, word embedding has the ability to express the semantic relationship between instructions.

\subsubsection{Position Embedding}

\label{pe}

In order for our model to perceive the order relationship between different instructions, we embed the absolute position $(1,...,m)$ of the  m instructions binary data as $p = (p_1,...,p_m)$, where $ p_j \in R^g $ is a column in a position embedding matrix $F \in R^{g \times S}$ ($S$ represents the number of input instructions ), indicating that the fixed position corresponding to the instruction is embedded as a g-dimensional vector, so that model can combine the position embedding and word embedding. Since the object file was not linked by compiler, the code segment data offset address starts from 0, so using absolute and relative addresses is equivalent. It is worth noting that the position embedding matrix is different from the embedding matrix mentioned in the previous section. It is a hard embedding representation. The matrix is constant, that is, the embedding representation corresponding to a certain position is fixed and is not trainable. The word embedding matrix mentioned in the previous section is trainable. For a given position $n$, where the dimension of the embedded representation is specified as $g$, the position $n$ is embedded as a vector $p_n$ as equation \ref{eq1} shows.

\begin{equation}	
p_n[k] = (1- \frac{n}{S+1} ) – (\frac{k}{g+1}) \times (1 – 2 \times \frac{n}{(S+1)} )
\label{eq1}
\end{equation}

Where $0<=k< g$, which represents the $k_{th}$ dimension of the position code. S represents the total number of instructions, $S = 1024$ in this paper.

Word embedding and position embedding both are combined to represent the input data  $e = (w_1+p_1,...,w_m+p_m)$ as the input data of the next layer. Position embedding representation is very useful in models, which gives the model the ability to capture instruction position information so that model can get higher accuracy.

\subsubsection{Convolution and Pooling}
\label{cnn}
We set all convolution kernel widths to be the same as the previous embedding layer output width and the lengths to $Conv\_K_1,Conv\_K_2,Conv\_K_3,Conv\_K_4$ (which means four different kinds of convolution kernels), and do not pad after convolution. 
After the convolution operation, we can get a $m-k+1$ dimension vector instead of a matrix (k for the convolution kernel length and m for the sequence length). 
Since we get the $m*n*h$ (h represents the distributed representation of each byte with the h-dimensional vector) after embedding the $m*n$ data (n represents the width of the input matrix). 
Then after convolution, we get a matrix of $(m-k+1) \times 1 \times 1 \times num\_filters$ ($num\_filters$ represents the number of convolution kernels per class),
Then use max pooling to take the maximum value of each channel to get a matrix of $1\times 1 \times num\_filters$.
Then splicing and reshaping the output of all convolution operations to get a $num\_filters \times filter\_sizes$ ($filter\_sizes$ represents the number of types of convolutional kernels) vector as input to the next layer of fully connected layers.

\begin{table*}[!hbt]
\caption{Experiment Results}
\begin{center}
\begin{tabular}{|c|c|c|c|c|c|c|c|}
\hline
\textbf{Model} &\multicolumn{5}{|c|}{\textbf{Precision with position embedding}}&\textbf{Model} & \textbf{Calculation} \\
\cline{2-6} 
\textbf{Structure} & \textbf{\textit{-O0}}& \textbf{\textit{-O1}}& \textbf{\textit{-O2/-O3}}& \textbf{\textit{-Os}}& \textbf{\textit{Accuracy}} & \textbf{Parameters} & \textbf{Speed}\footnotemark[1] \\
\hline
\textbf{\textit{BinEye}} & 0.9800 &\textbf{\textit{0.9690}} &\textbf{\textit{0.9807 }} &\textbf{\textit{0.9606}} & \textbf{\textit{0.9724}} &306439 & \textbf{\textit{65.29}} \\
\hline
LSTM&0.9805& 0.9577&0.9525&0.9367 &0.9408 & 101700 &8.79 \\
\hline
GRU& \textbf{\textit{0.9839}}& 0.9464& 0.9391& 0.9587 &0.9507 & \textbf{\textit{95492}} & 7.37 \\
\hline
CNN+LSTM  & 0.9637 &0.7699  &0.7457 & 0.4757 &0.7784 & 4349060 & 5.53\\
\hline
CNN+GRU  & 0.9794& 0.9464& 0.8588& 0.8966 &0.9391 &3298436&2.43\\
\hline
\end{tabular}
\label{tab2}
\end{center}
\end{table*}

\begin{figure}[htbp]
    \centerline{\includegraphics[width=3.5in]{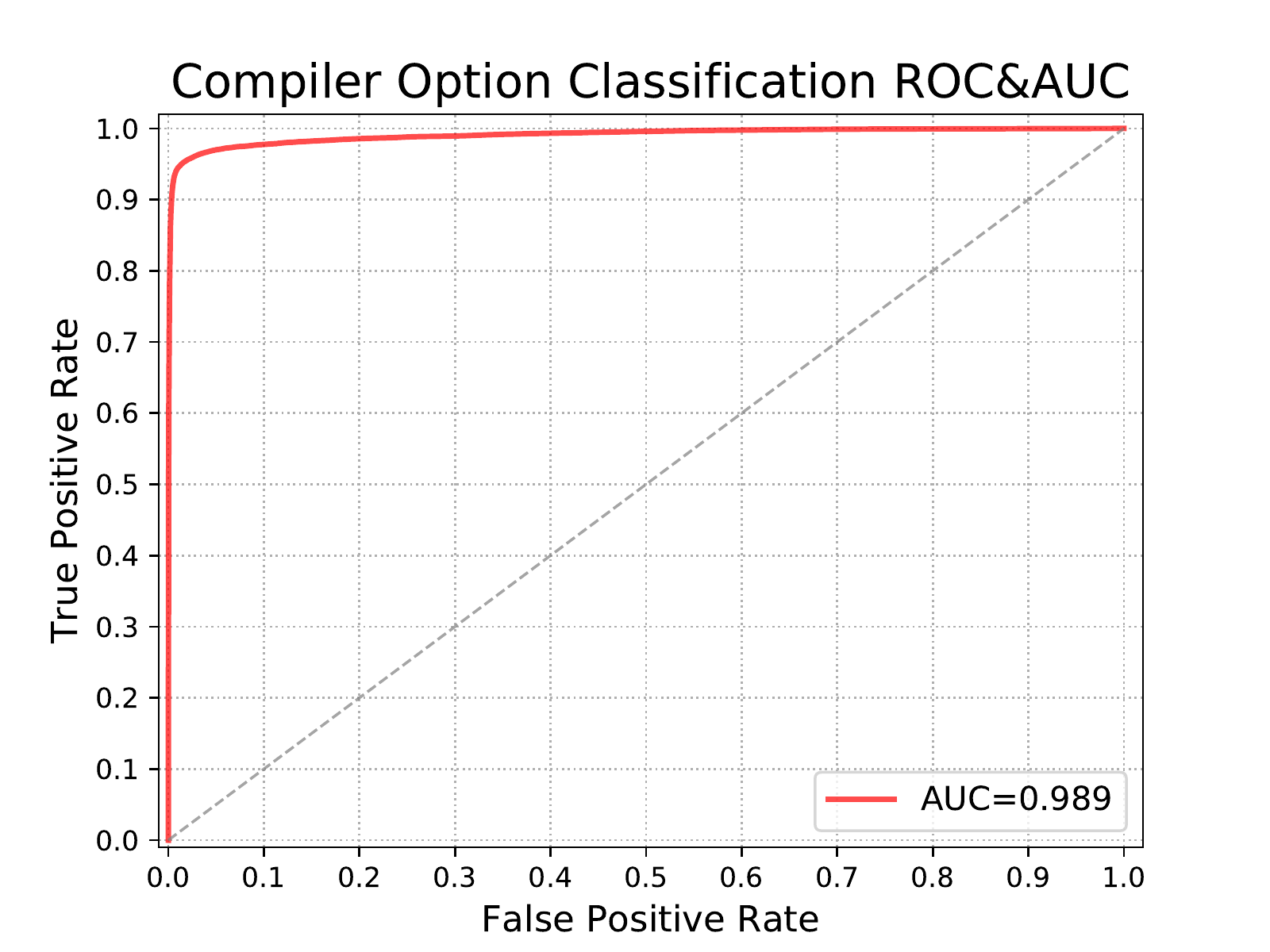}}
    \caption{ROC of Compiler Optimization Level Classification.}
    \label{roc}
\end{figure}

\subsubsection{Output}

The last part of the model is the fully connected layer, which enhances the expressive power of the model. The output of the fully connected layer uses the softmax function to obtain the classification probability for each class, as defined below equation \ref{softmax}

\begin{equation}
S_i = \frac{e^{V_i}}{\sum_{i}^{C}e^{V_i}}
\label{softmax}
\end{equation}

Where $V_i$ is the ith value of the dense layer output vector, and C is the total number of classes. Since we have -O0, -O1, -O2/-O3, -Os four classes in total, C has a value of 4. Because the output of the model is a 4-dimensional vector, where i-th dimension represents the probability of the ith class. we use argmax function to get the subscript with the largest probability dimension as the predicted class.

\section{Experiment}
This part mainly describes the training environment of the model, the training and test results of the model, and the analysis of the experimental results.

\footnotetext[1]{The calculation speed is expressed as the number of calculations per second}
\subsection{Model Performance}

Our model \textbf{BinEye} is fully based on CNN and equipped with position embedding.  Due to the weight sharing characteristics of CNN,  the total number of parameters of our neural network model is 30, 639,  and the calculation of 457 samples in the forward direction takes only 7 seconds which is much faster than RNN model. Our model has achieved a 97.24\% accuracy rate which is more accurate than other RNN model.


Sometimes the order of the program statements is very important for the control flow,  which requires our model to have the ability to capture the order of sequences. In this paper,  by introducing positional embedding,  each statement position is numbered,  and each number corresponds to a vector.  By combining the position vector and the instruction embedding vector,  a certain position information is introduced for each instruction. 
We do an experiment on the model with the ability to identify different compiler optimization. In order to better enable the model to capture the semantic differences of instruction combinations of different lengths, we have tried different length convolution kernels and different combinations, and finally obtained a better performance convolution kernel combination.
. Four different types of convolutional kernels are used in the experimental model. The size of each convolution kernel is only different in length, which is 2,3,4 and 5 respectively. Our model took a few hours to train on the training dataset and eventually got an accuracy on the test dataset. According to Table \ref{tab2}, we can know that the recognition accuracy of the model \textbf{BinEye} with kernel size 2,3,4,5 for each compiler optimization levels -O0, -O1, -Os, -O2/-O3 is above 95\%, accuracy for four kinds of compiler optimization levels reached 97.24\%.

\subsection{Compared With RNN Model}
For comparison, we have built several RNN neural networks. We also use word embedding representation technology in the RNN neural network. After word embedding layer, we built RNN layer which is one of \textbf{LSTM, GRU, CNN+LSTM, CNN+GRU}. Then we built a dense layer as the output layer.
we use the same dataset to feed RNN model we built, and the RNN model took a long time to train and test on the same machine as \textbf{BinEye} model. Then we get the four type RNN network test result for comparison with our CNN-based model, at the same time we record the model size (Number of parameters) and the forward calculation speed (The number of calculations per second). The comparison result is shown in Table \ref{tab2}. we can learn that our model is far superior in terms of accuracy and speed of calculation.




\subsection{Application on Object Files}
 For the reason that we only cared about 4K size code section which belongs to a part of the code segment of object file during model training and testing, the final accuracy rate of 97.24\% may not fully represent the BinEye's ability to recognize the optimization levels for the object file. So we reloaded the trained model and test the ability to recognize the optimization levels for the object file in our built dataset and we got a final accuracy of 97.49\% on the whole dataset. For one single object file, we split it into several 4K size binary code blocks. We used BinEye to recognize the optimization level of every single code blocks, then we take the mode of the recognition results of several code blocks as the final result of the object file. We counted the TPR and FPR under different thresholds and got the ROC curve as shown in Fig \ref{roc}.

\begin{figure}[htb]
    \centerline{\includegraphics[width=3.5in]{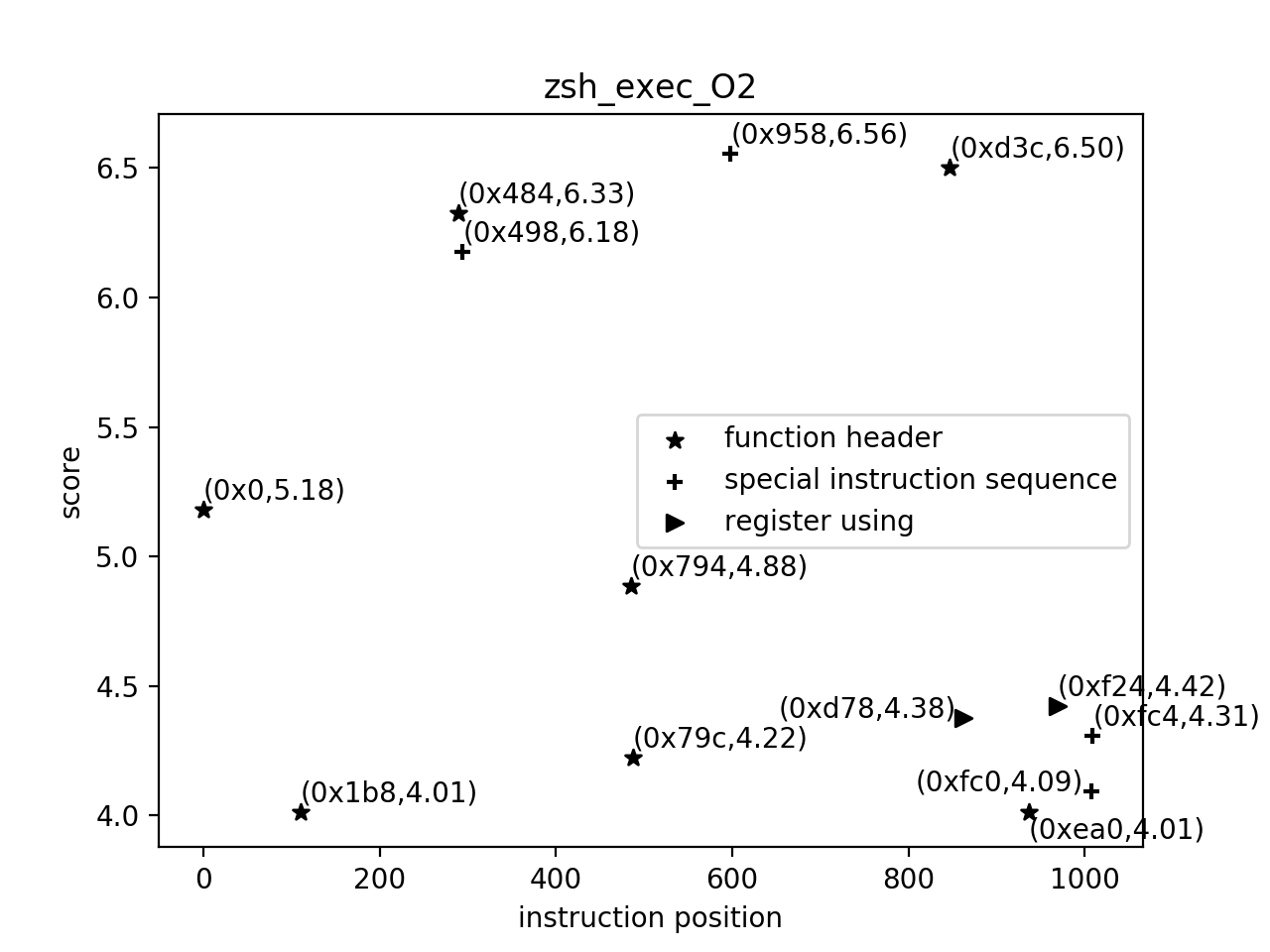}}
    \caption{Convolution and Pooling output}
    \label{o2score}
\end{figure}

\section{A Method to Analyze Compilation Optimization}

\label{analysis}

We analyzed the outputs of convolution and pooling part and found that only a few outputs are the non-zero value, which means not all instructions have a contribution to compilation optimization level recognition task. We started to focus on the instruction sequences corresponding to non-zero outputs. After analyzing dozens of object cases, we took two typical file exec\_O0.o and exec\_O2.o which were compiled from same source code and by different optimization level for analysis. The two files can be obtained from github \cite{github}.

We got a few large output values from the outputs of convolution and pooling operation for exec\_O2.o since the large value has a greater impact on the final classification. Figure \ref{o2score} shows the outputs that we got. The comment for each point represents the score of corresponding position code block at one address. The x-axis represents the position of the instruction in the 1024 instructions of model input. By analyzing the higher-score instructions and comparing them with the instruction sequences at the corresponding locations of the exec\_O0.o file, we have proposed a method to analyze the difference between the homology binary code with different compilation optimization levels. The method steps are as follows.

\begin{enumerate}
\item Extract the binary data of code segments and divide the binary data into 4K sizes.
\item Input the code segment data into the BinEye model, perform forward calculation, and obtain the results after convolution and pooling.
\item Obtain the corresponding instruction address whose output value is greater than 0, and then extract consecutive 4 instructions backwards.
\item Summarize the patterns corresponding to successive instructions, and compare other code with different compilation optimization levels of the same source code.
\item Compare the differences in patterns with other compiler optimization levels codes, and summarize the differences in different levels of compilation optimization.
\end{enumerate}
Through our method, we found that there are three types of differences between the two different compilation optimization levels, which are all marked in different shapes in the Figure \ref{o2score}.  For the three different differences: function headers, special instruction sequences, register usage, we'll cover them in the following sections.

\begin{figure}[htb]
    \centerline{\includegraphics[page=1,width=3.5in]{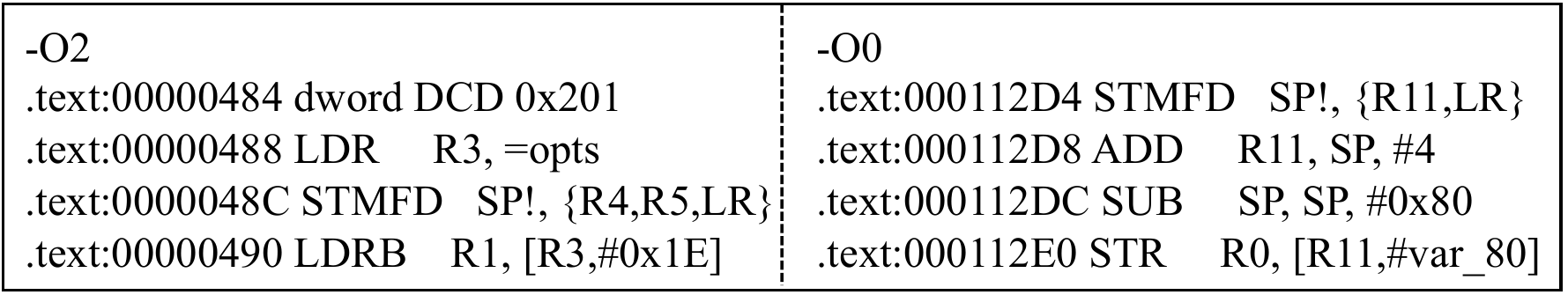}}
    \caption{instructions in 0x484 of -O2 and corresponding instructions of -O0 for function header analysis}
    \label{funcheader}
\end{figure}

\subsection{Difference in Function Header}

According to Fig \ref{o2score}, we first analyzed the instruction sequence corresponding to \textbf{(0x484, 6.33)}. Corresponding we also analyzed and compared the codes of the corresponding function header of the -O0 object file exec\_O0.o. We got a very clear difference between -O0 and -O2 compilation optimization level. We found that code compiled with -O0 level always saves R11 register into the stack on the function header and restore it when function exists. But in codes compiled with -O2 level, One or more of R4-R10 registers were used at the function header by instruction STMFD. The difference sample is shown in Fig \ref{funcheader} (-O2 uses instruction \emph{STMFD SP!,\{R4,R5,LR\}}, while -O0 uses \emph{STMFD SP!,\{R11,LR\}}). By analyzing the difference in function headers between the two optimization object files in a large amount (such as (0xd3c,6.50),(0x0,5.18)), we think this can be a good basis for distinguishing different compiler optimization level. It is very impressive that BinEye can recognize the nuances of register usage in the function header.

\begin{figure}[htb]
    \centerline{\includegraphics[page=2,width=3.5in]{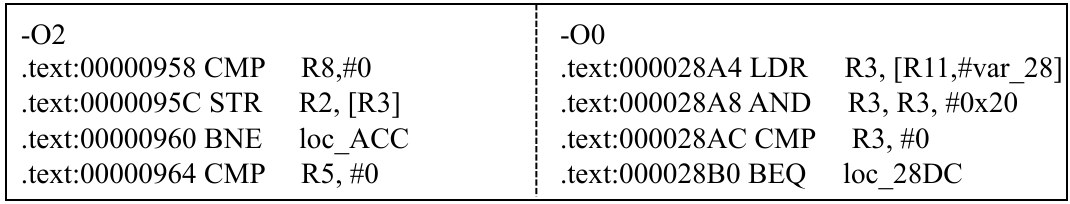}}
    \caption{instructions in 0x958 of -O2 and corresponding instructions of -O0 for special instruction sequence analysis}
    \label{instruseq}
\end{figure}

\subsection{Special Instruction Sequence}

We analyze the instruction sequence of length 4 according to \textbf{(0x958, 6.56)} shown in Fig\ref{o2score} and corresponding instruction sequence in -O0 object file. Then we compare similar instruction combinations elsewhere in the file (such as (0x498,6.18)). We found a very interesting combination of instructions. Instruction combination "CMP BEQ" or "CMP CMPEQ BEQ" often appear together under -O0 optimization level, but under -O2 optimization level, instruction combination "CMP BNE" or "CMP CMPEQ BNE" often appear together. The code sample is as shown in the Figure \ref{instruseq}. 

\begin{figure}[htb]
    \centerline{\includegraphics[page=3,width=3.5in]{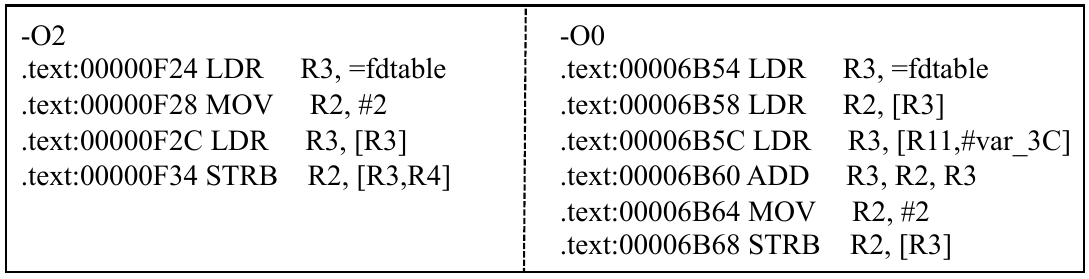}}
    \caption{instructions in 0xF24 of -O2 and corresponding instructions of -O0 for register usage analysis}
    \label{reguse}
\end{figure}

\subsection{Register Usage}

Register using is an important rule in ARM Architecture Procedure Call Standard\cite{armdoc}. By analyzing the instruction sequence of length 4 according to \textbf{(0xf24, 4.42)} shown in Figure \ref{o2score} and corresponding instruction sequence in -O0 object file. We found that code under optimization level -O2 use registers as much as possible to store local variables. But under -O0 optimization level, the code always pushes the local variables into the function stack, and load them when needed. As shown in Figure \ref{reguse}, code under  optimization level -O0 use "LDR R3, [R11,\#var\_3C]" load a local variable, and save immediate $2$ into memory address corresponding "parameter+fdtable". But in code under  optimization level -O2, register R4 was used to hold the local variable.

By analyzing the outputs of the BinEye convolution and pooling part and comparing the instruction sequences of the corresponding parts of optimization level -O0 and  optimization level -O2, we did find some unique patterns between different compilation optimization levels. That is to say, BinEye remembers some special instruction patterns related to the compilation optimization by continuously learning the input binary code so that BinEye can recognize the compilation optimization levels of different object files.

\section{Conclusion}
This paper mainly proposes an efficient deep learning model without any prior knowledge: \textbf{BinEye}, which is used to identify the compilation optimization levels of the object file.  Model based entirely on CNN,  with fewer parameters, high forward calculation speed and fast training speed,  etc. We did a lot of preparation work first,  specifying five different compilation optimization levels to compile hundreds of open source components and generating a large dataset for model training and testing. According to the classification accuracy of the model,  we continuously adjust the parameters and modify the network structure.  Finally,  the classification task accuracy of the model is about 97.24\%,  which achieves an ideal effect.  Through the analysis of the model output, we found that the model can memorize the special instruction patterns between different compilation optimization levels.
With the help of the BinEye, we successfully analyzed and verified several differences in assembly code at different levels of compilation optimization.

\section{Acknowledgement}
The research was supported by National Natural Science Foundation of China (No.U1636120), Strategic Priority Research Program of Chinese Academy of Sciences (No. XDC02020100) and Key Program of National Natural Science Foundation of China (No.U1766215).

\end{document}